# Polariton condensates at room temperature

T. Guillet, C. Brimont

Laboratoire Charles Coulomb (L2C), UMR 5221, Centre National de la Recherche Scientifique (CNRS)-Université de Montpellier, Montpellier, F-FRANCE

**Abstract**

We review the recent developments of the polariton physics in microcavities featuring the exciton-photon strong coupling at room-temperature, and leading to the achievement of room-temperature polariton condensates. Such cavities embed active layers with robust excitons that present a large binding energy and a large oscillator strength, i.e. wide bandgap inorganic or organic semiconductors, or organic molecules. These various systems are compared, in terms of figures of merit and of common features related to their strong oscillator strength. The various demonstrations of polariton laser are compared, as well as their condensation phase diagrams. The room-temperature operation indeed allows a detailed investigation of the thermodynamic and out-of-equilibrium regimes of the condensation process. The crucial role of the spatial dynamics of the condensate formation is discussed, as well as the debated issue of the mechanism of stimulated relaxation from the reservoir to the condensate under non-resonant excitation. Finally the prospects of polariton devices are presented.

In the past decade, the semiconductor microcavities have been widely used for both the fundamental studies of the polariton collective states such as condensates and quantum fluids, and the demonstrations of original quantum optoelectronic devices: polariton lasers, switches, transistors. For both topics, the prospect of an operation at room temperature has raised a strong interest and led to the development of specific microcavities based on new active layers. The key requirements for strong coupling up to room temperature are a large binding energy and a large oscillator strength of the excitons; a broad spectrum of active layers is presently being explored among inorganic semiconductors (mainly GaN, ZnO, ZnSe, CuCl, CuBr) and organic molecules (dyes, J-aggregates, perovskites). Their integration into optical cavities jointly requires a strong technological effort, so to reach quality factors (Q) typically larger than 1000. Some of those room-temperature microcavities are now mature enough to allow in-depth fundamental studies of the polariton condensates, their formation, propagation and control, as well as examples of polariton devices. We give here an overview of the recent demonstrations achieved with such microcavities, and emphasize their specificities compared to GaAs and CdTe-based microcavities.

1) Robust excitons for room temperature polaritons

In the choice of the appropriate active layer, the main criterium for a room temperature operation of a microcavity is the following: the exciton binding energy and the polariton Rabi splitting should be both larger than the thermal energy at 300 K ($k_B T$ = 26 meV). The search for the strong exciton-photon coupling at room temperature therefore started with organic molecules in 1998 [1] and wide bandgap semiconductors, especially GaN in 2003 [2–4]. ZnO [5–9] and CuCl [10] rapidly emerged as alternative candidates for strong coupling, up to T = 550K in ZnO [11,12], as well as ZnSe [13–16] and CuBr [17]. Organic active layers include perovskites [18,19], J-aggregates [1,20], anthracene [21], fluorene (TDAF) [22], conjugated polymers [23]. As detailed in the Table 1, all those materials exhibit an exciton binding energy larger than 26 meV, as well as a longitudinal-transverse splitting larger than a few meV (compared to 80 µeV for GaAs), leading to large Rabi splittings. It should be noticed that bulk active layers and quantum wells are both deeply investigated as active layers.

| Active layer | Cavity geometry | Exc. energy (300K) | Exc. binding energy | Max Cavity Q | Polariton Rabi splitting | Strong coupling | Polariton laser | Condensation phase diagram | Condensation models | Additional demonstrations |
|---|---|---|---|---|---|---|---|---|---|---|
| **Inorganic** | | | | | | | | | | |
| GaN bulk | Planar microcavity | 3.42 eV | 25 meV | 2800 | 20-60 meV | [2–4] | [24] | [25] | [26] | Spatial coherence [24]; Spontaneous polarization buildup [27] ; electrically injected polariton laser [28] |
| | Nanowire WGMs | 3.42 eV | 25 meV | 700 | 115 meV | [29] | | | | |
| GaN QWs | Planar microcavity | 3 - 3.3 eV | 25 meV | | 20-60 meV | [30] | [31] | [32] | [32] | Electrically injected polariton diode [33] |
| ZnO bulk | Planar microcavity | 3.31 eV | 60 meV | 2600 | 50-250 meV | [5,7,8,34] | [35,36] | [25,37,38] | [37,39,40] | Condensate propagation [38,41] ; Condensate localization [42] |
| | Nanowire FP | 3.31 eV | 60 meV | 260 | 100 meV | [43,44] | | | | |
| | Nanowire WGMs | 3.31 eV | 60 meV | 800-1200 | 50-310 meV | [11,12,45,46] up to 550K | [47,48] up to 450K | | | Parametric scattering [49]; 97% excitonic condensates, spatial coherence [50]; Parametric relaxation [51]; 1D polariton superlattice [52] |
| ZnO QWs | Planar microcavity | 3.35 eV | 80 meV | 300 | 26 meV (5-200K) | [9] | | | | |
| ZnSe QWs | Planar microcavity | 2.31 eV | 40 meV | 3850 | 20-40 meV | [13] (175K); [14,15] (300K) | | | | Polariton refrigerator [53] |
| | Micropillar | 2.31 eV | 40 meV | 5000 | 32 meV | | [16] (6K) | | | |
| CuCl bulk | Planar microcavity | 3.2 eV | 190 meV | 250 | 100-160 meV | [10] (10K), [54] (300K) | | | [55] | |
| CuBr | Planar microcavity | 3 eV | 108 meV | 490 | 28-100 meV | [17] | [56] | | | |
| **Organic** | | | | | | | | | | |
| Perovskites | Planar microcavity | 2.4 eV | 200 meV [19] | 25-100 | 140-230 meV | [18,57] | | | | In-plane confinement, Q=750 [58] |
| J-Aggregates | Planar microcavity | 1.8 eV | 360 meV | 150 | 80-160 meV | [1,20] | | | [59–61] | |
| Anthracene | Planar microcavity | 3.16 eV | | 800 | 50-100 meV | [21] | [62] | | | |
| Fluorene (TDAF) | Planar microcavity | 3.5 eV | | 600 | 600 meV | [22] | [22] | | | Spatial coherence [63] |

| | | | | | | |
|---|---|---|---|---|---|---|
| Conjugated polymer | Planar microcavity | 2.7 eV | >700 | 120 meV | [23] | Spatial coherence [23] |

**Table 1:** Comparison of the room-temperature polariton systems, and the corresponding active layers and optical resonators. Perovskites are hybrid organic-inorganic molecular crystals, whereas the other organic systems consist of active layers of small molecules.

Once the active layer is inserted inside a microcavity, the Rabi splitting deduced from the mode anticrossing assesses the strong coupling regime between excitons and photons (see Table 1), as probed by reflectivity or transmission measurements. The main impediment towards polariton luminescence, polariton lasing or condensation is the improvement of the microcavity quality factor on the one hand, and the radiative efficiency of free excitons at 300K on the other hand. For each material, this requires an intense technological effort in order to maintain an excellent crystalline quality of the active layer while improving the cavity features. For GaN-based microcavities, this led to the development of crack-free lattice-matched nitride distributed Bragg reflectors (DBRs) [64], and more recently crack-free nitride DBRs on patterned silicon substrates [65]. For other materials, a complex combination of nitride DBRs and dielectric DBRs is often implemented, while preserving the quality of the active layer [57,66].

While most works have been reported with planar microcavities, alternative optical resonators are also explored. The possibility of a self-organized growth of GaN and ZnO into hexagonal nanowires [43,67,68] and microwires [11,12,29,45,46,69] has led to interesting resonators which crystalline quality is improved compared to planar structures. The exciton then couples to either longitudinal Fabry-Perot modes in cleaved nanowires (like in a finite length monomode optical fiber), or whispering gallery modes in micron-thick wires. The strong coupling has been demonstrated, with larger Rabi splittings than in planar microcavities, due to the optimal overlap between the photon mode and the free exciton (see Table 1). Moreover this offers the possibility to explore 1D polaritonic systems without requiring the implementation of the complex lithography and etching processes presently used to obtain 1D polaritons in GaAs wire-shaped cavities [70].

As discussed in the following sections, the achievement of a good photon confinement in the resonators appears as a requirement for the further demonstrations of polariton lasers. The minimum quality factor of the photonic modes is of the order of 1000 in the case of GaN and ZnO active layers, and diminishes to 500 for CuBr and 200 for anthracene, both with larger exciton binding energies.

2) Impact of the large oscillator strength on the polariton eigenmodes

The impact of the large oscillator strength of the excitonic transitions has been studied in details in the case of ZnO. In early works on ZnO nanowires [43,44], the deviation between the Fabry-Perot free spectral range far and near the excitonic transitions is discussed as a first signature of the strong coupling between excitons and confined photons. Indeed the refractive index of ZnO presents pronounced resonances corresponding to the exciton states as shown on Figure 1.a, so that the textbook resonance condition for the P$^{th}$ Fabry-Perot resonance $n_{ZnO}(\lambda) d_{ZnO} = P\lambda/2$ is fulfilled for a few values of P. Let us compare the cases of two planar microcavities embedding a bulk GaAs and a bulk ZnO active layer. In the GaAs microcavity [71], the cavity features are understood through the introduction of the GaAs refractive index (real part) in the electromagnetic modelling of the cavity, whereas the coupling with excitons is described by a Rabi splitting proportional to the GaAs exciton resonant absorption. Such a separate treatment is possible because the excitonic resonance induces a change of the coupled mode energy, but the electromagnetic field distribution remains almost unchanged. This is not at all the case in the bulk ZnO microcavity, nor with GaN or organic active layers, because both the real and imaginary parts of the dielectric function are strongly varying at the excitonic resonance. A bulk ZnO microcavity in the strong coupling regime therefore presents a few lower polariton branches (LPBs) within a narrow spectral range (Figure 1.b), corresponding to different photonic modes, in contrast to GaAs or CdTe microcavities; their modeling is complex and still debated [72,73]. As an example, the figure 1.c presents the simulated angular-resolved reflectivity of a ZnO microcavity in the case of a 5λ/4 mode at almost zero detuning with the exciton energy; an additional 7λ/4 mode is observed simultaneously.

The second consequence of the large oscillator strength concerns the damping of the upper polariton branch (UPB) due to the absorption of the electron-hole continuum. When the Rabi splitting is larger than the excitonic binding energy, the UPB is indeed resonant with this continuum, so that the corresponding transition is strongly broadened and can hardly be identified in the reflectivity and emission spectra: the reflectivity spectra present a very broad dip above the exciton energy (at 3.41-3.44 eV in the figure 1.c), much broader than the LPB mode [34,74]. In a ZnO microcavity with a cavity length larger than λ, only the LPBs are well-defined states of the system.

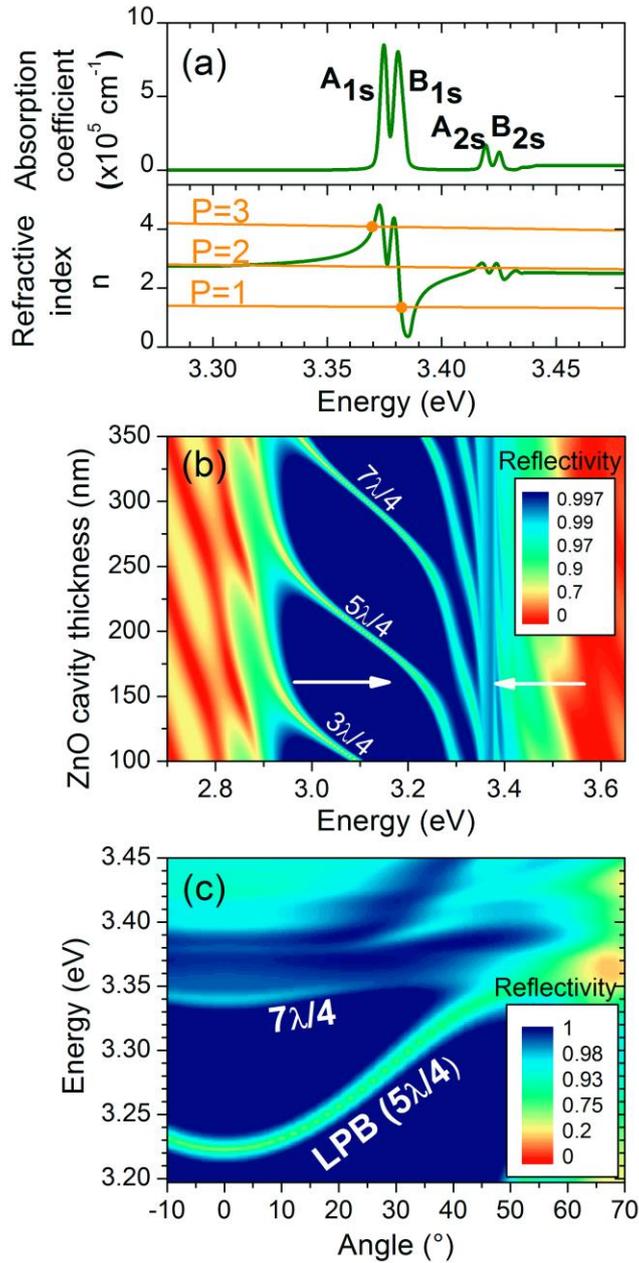

**Figure 1:** (a) Absorption coefficient and refractive index of ZnO for a 5 meV homogeneous broadening [75–77]; the resonance condition for cavity modes with different interference orders P is shown as lines. (b) Dependence of the reflectivity spectra on the thickness of the ZnO active layer, obtained from transfer matrix simulations under normal incidence. The cavity consists in a ZnO active layer (MBE) embedded between a bottom (Ga,Al)N/AlN DBR (MBE) and a top (Si,O)/(Si,N) DBR (PECVD), on a Si(111) substrate [34,35,78]. (c) Angular resolved reflectivity simulated for the cavity thickness indicated by a horizontal arrow on (b). From Ref. [79].

A third consequence of the large Rabi splitting is the enhanced temporal coherence of the polariton states at high temperature, compared to the bare exciton state. At room temperature, the polariton spectral linewidth at negative detuning is indeed much smaller than the one deduced from the weighted average of the photon and exciton bare linewidths [46]. This is interpreted as a spectral protection from exciton-phonon scattering processes when the energy difference between the LPB and the exciton reservoir is larger than the LO phonon energy.

3) Demonstrations of polariton lasers at room temperature

Most room-temperature polariton lasers operate under a pulsed and focused non-resonant excitation that is necessary in order to reach a large exciton density in the reservoir. As an example, such a demonstration is presented for a bulk ZnO microcavity at almost zero detuning and 300K in the figure 2 [37]. As the excitation density is increased, a non-linear increase of the LPB emission is observed, together with a reduction of its spectral linewidth. This is accompanied by a blueshift of the LPB emission energy that reflects the interactions of the condensate with the exciton reservoir. The strong coupling is maintained across threshold if this blueshift is smaller than the Rabi splitting of the cavity; this condition is properly fulfilled in the case shown on figure 2.

The first room temperature polariton laser was demonstrated in a bulk GaN microcavity [24,27]. It was later reached with GaN QWs [31], bulk ZnO in planar cavities [36–38] as well as ZnO nanowire WGMs [47,48], anthracene, fluorene and conjugated polymers in a planar cavity [22,23,62] and more recently CuBr [56]. In each case, these demonstrations resulted from a progressive improvement of the photonic quality factor while preserving a high quality active layer with a proper radiative efficiency up to 300K and large carrier densities. From the overview presented in table 1, it appears that the minimal quality factor allowing for polariton lasing decreases as the exciton binding energy and oscillator strength of the active layer increase, with a typical value of 500-1000 in the cases of GaN and ZnO, 500 for CuBr and eventually 200 for anthracene.

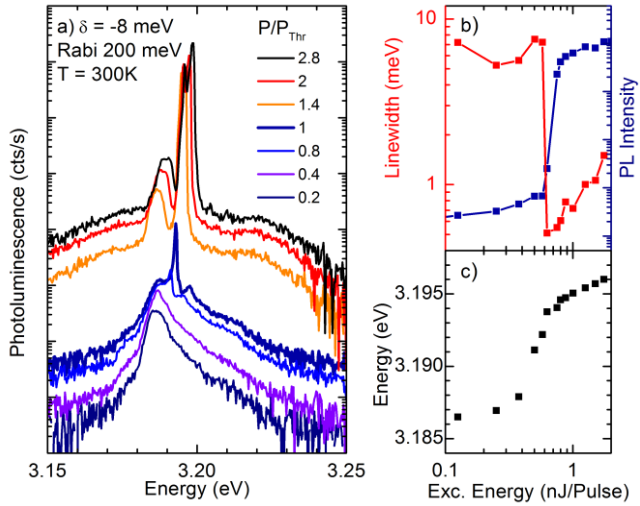

Figure 2: Polariton lasing in a bulk ZnO microcavity at almost zero detuning at 300 K (2.5 λ cavity at δ=-8 meV, for a Rabi splitting of 230 meV). (a) Power dependent series of angle integrated micro-PL spectra. The polariton lasing threshold ($P_{thr}$ = 0.6 nJ/pulse) corresponds to the thick blue spectrum. The excitation power density relatively to the threshold power is indicated for each spectrum. Note that the structure visible beyond threshold near 3.185 eV is only an artefact (ghost) of the spectrometer. (b) Integrated intensity, linewidth and (c) energy of the LPB transition as a function of the excitation power density. From Refs. [37,66].

4) Tunable polariton condensates: the condensation phase diagram

Beyond the demonstration of polariton lasing at room temperature, a further step forward lies in the ability to tune the exciton-photon fraction of the condensate, and to follow the temperature dependence of the condensation dynamics. This allows a detailed comparison with the kinetic and thermodynamic models for condensation. This requires (i) the existence of a thickness gradient in the photonic resonator, so to tune the exciton-photon detuning δ, and (ii) a good homogeneity of the active layer, such that the threshold for polariton lasing mainly depends on the detuning rather than on extrinsic sample inhomogeneities. The large exciton oscillator strength of the active layers allows for a wide tunability of the exciton-photon content: in the case of ZnO, exciton fractions ranging from 17% to 96% have been demonstrated for the polariton condensates in planar microcavities [37], and 97% in ZnO microwires [50].

The first experimental investigation of the condensation threshold as a function of the temperature and the detuning has been performed on a nitride microcavity embedding GaN/AlGaN multiple quantum wells (Figure 8 in Ref. [32]). This work provided the opportunity to test the kinetic model based on Boltzmann rate equations developed for 2D infinite systems [26,39,80]. A similar study performed on a ZnO microcavity [37] is presented in the figure 3. Both works provide evidence of two distinct regimes of condensation dynamics. At zero and positive detunings, polaritons are thermalized due to an efficient relaxation and a proper polariton radiative lifetime; the condensation occurs in a thermodynamic regime (II). At negative detuning, the distribution of the polariton occupancy along the LPB is strongly out of equilibrium due to the small excitonic content of the polariton branch and the large energy difference between the condensate and the excitons; the condensation is kinetically driven (I) and the polariton laser behaves like a conventional laser. This competition between relaxation and polariton losses was already identified in early works on GaN and ZnO microcavities [6,81] in which the so-called bottleneck effect in the polariton relaxation prevented any observation of polariton lasing due to the insufficient quality factor of the cavities.

The large temperature range for polariton laser operation offers the possibility to explore finely the transition between the thermodynamic and kinetic regimes. The thermodynamic regime is only obtained for strong positive detunings ($\delta$ comparable to the Rabi splitting) at 5K due to the poor efficiency of the phonon-mediated relaxation, whereas it extends to zero detuning at 300K. In both GaN and ZnO systems, the minimal threshold, i.e. the minimal exciton reservoir density at condensation, is obtained at the frontier between the two regimes [25]. Moreover, the large tunability of the exciton-photon content of the obtained polariton condensates allows controlling the interaction between polaritons (i.e. the exciton fraction of the polaritons), which is crucial for the further developments of nonlinear functionalities.

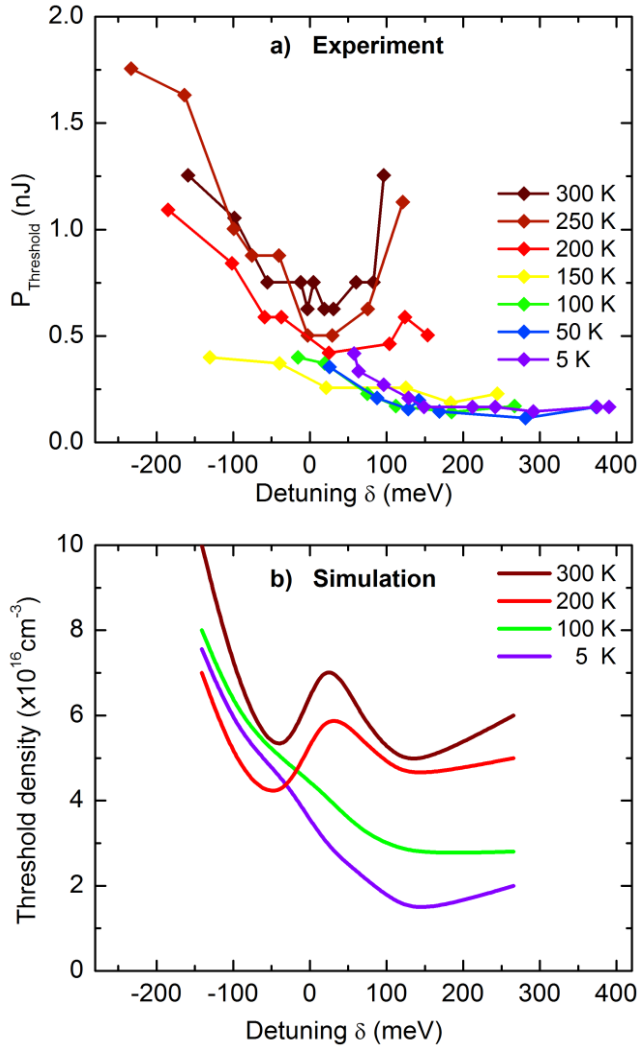

**Figure 3**: (a) Experimental threshold for polariton condensation in a 2.5λ bulk ZnO microcavity (Q=2600, Rabi splitting 200 meV), from Ref. [36]; two condensation regimes are identified (I – out of equilibrium, driven by the kinetics; II – thermodynamics-like) (b) Simulation using the semi-classical Boltzmann equation, from Ref. [36]; the two minima are related to a resonant enhancement of the LO-phonon relaxation.

It should be noticed that the crossover from out-of-equilibrium to thermalization can also be observed in a microcavity operating in the weak coupling regime, and does not require polaritons. If a photon gas is brought into equilibrium with the excitations of dye molecules through a very efficient emission/reabsorption process, the transition from a laser regime to a photon Bose-Einstein condensation can be observed at room temperature [82,83].

The spatial dynamics of the polariton condensation process is an important issue. Indeed most investigations are performed under a tightly focused excitation, with a laser spot diameter of the order of 1-5 µm that is comparable to the typical distance for a ballistic propagation of the polaritons. As evidenced on figure 2, the polariton condensate presents a slight blueshift at threshold (8 meV in this figure, i.e. 4% of the Rabi splitting) that is a signature of the Coulomb repulsion of the exciton reservoir. As theoretically investigated by M. Wouters *et al.* [84], the polariton condensate nucleates under the excitation spot and it is propelled outwards by this repulsive potential, so that the condensation threshold is not defined anymore as the balance between the polariton loss rate and the stimulated relaxation rate from the reservoir to the condensate: the outwards flux of polaritons within the condensate also has to be compensated, and it dominates the polariton loss rate. In a recent study performed in a ZnO microcavity with a low photonic disorder, a typical 10-fold increase of the condensation threshold is extracted from the modeling of the condensate imaging and its ballistic propagation [41]. In most works on room-temperature polariton lasing, this threshold increase related to the tightly focused excitation regime therefore plays a very important role [36,38,48–50,63,85,86]. A precise investigation of the threshold excitation density as a function of the excitation spot diameter has shown that the optimal diameter of the excitation is of the order of 10 µm, for very similar GaN and ZnO microcavities [25]. The comparison between small and large spot excitations has also been explored at room temperature in the case of TDAF organic microcavities, up to 100 µm spot diameters [22,63]. The precise shape of the excitation spot (for example gaussian or flat-top) has a strong impact on the condensate wavefunction [63,87,88]. A second signature of the spatially out-of-equilibrium nature of the polariton condensate lies in the large contribution of k≠0 wave-vectors in the condensate wavefunction, as shown in the Fourier-space imaging of the condensate [38,41,89]. Finally, the photonic disorder [90] may strongly modify this first interpretative picture, and lead to a complex interplay between polariton localization, propagation and radiative losses [42], like in CdTe microcavities [91]. The demonstration of the spatial coherence of the condensates is therefore complex to establish and to interpret [42,50,63].

5) Stimulated relaxation mechanisms: a debated issue

Under non-resonant excitation, the relaxation from the exciton reservoir to the condensate is a complex issue. The first invoked relaxation processes are the exciton-exciton Coulomb scattering and the exciton-LO phonon coupling. Their respective contributions, and especially the observation of minima of the condensation threshold at the energies $E_X$-$E_{LO}$ and $E_X$-$2E_{LO}$ are properly modeled in the kinetic model presented in the previous section (Figure 3.c) [25,32,37,85] ($E_X$ and $E_{LO}$ are the bare exciton energy and the LO phonon energy). However the biexciton formation and energy transfer to the polariton branch is also proposed as an efficient process feeding the condensate if the condensate and the $X_2$ states are brought into resonance [50,92]. Even if the order of magnitude of those three processes is very different, this relaxation picture is very similar to the one of GaAs or CdTe microcavities. They provide an efficient scattering mechanism for parametric processes along the polariton branches [49,51].

More specifically, ZnO (as well as II-VI semiconductors and possibly other compounds with large exciton binding energy) presents an additional relaxation process called the P-band. At large exciton density, but below the Mott density, two excitons can scatter through the Coulomb interaction; the final state is an exciton in an excited state and a photon, instead of two excitons, fulfilling the energy and momentum conservation rules like the standard exciton-exciton scattering process. The energy of the emitted photon is about one effective Rydberg ($R_y^*$) below the exciton energy (i.e. ~$E_{gap}$ − 2 $R_y^*$, where $E_{gap}$ is the bandgap of ZnO). This process provides an efficient gain band, as studied in the 1970's in II-VI bulk semiconductor lasers [93,94], and in the 2000's in ZnO nanorod lasers [44,95–97]. In the planar microcavity geometry at moderate Q-factor, it contributes to the optical gain when operating the microcavity in the weak coupling regime; the ZnO microcavity here operates as an optically pumped vertical cavity surface emitting laser (an exciton-VCSEL) below the Mott transition when the bare cavity mode is resonant with the P-band [78], at negative detuning ($\delta$=-120 meV, Rabi splitting 130 meV). It is worth mentioning that the same microcavity operates as a polariton laser in the strong coupling regime at T=120K at a smaller negative detuning ($\delta$=-20 meV) [35], as also recently discussed in ref. [98]. This highlights the complex competition between scattering processes and lasing mechanisms.

Four mechanisms therefore compete and provide efficient channels of stimulated relaxation in ZnO microcavities: LO-phonon relaxation, exciton-exciton Coulomb scattering, resonant transfer through the biexciton state, and P-band scattering. Their relative contributions depend on the exciton-photon strength (the Rabi splitting), the polariton lifetime, the detuning controlling the resonances with LO-phonon replicas of the exciton or the biexciton transition, and the reservoir density. Finally the possible exciton saturation and/or exciton ionization into an electron-hole plasma enters into play at large polariton densities, beyond the condensation threshold [99].

6) Towards 300K polariton devices

The achievement of polariton condensation at room temperature with various excitonic active layers is a promising first step for the development of more advanced polariton demonstrators and devices. The first perspective concerns the electrical injection of a polariton laser. After the early theoretical proposal [100] and patent [PCT/FR2009/001257], and the preliminary demonstration of a GaN polariton diode [33], a first attempt has been published recently in the case of a GaN microcavity, with a complex in-plane geometry [28]; its transfer to a vertical cavity geometry with a larger Q factor remains challenging. The possibility to inject or amplify waveguide polaritons below the light cone has been recently proposed [40].

Beyond the realization of an electrically-injected polariton laser operating at 300K, a broad spectrum of more prospective polariton devices have been proposed during the last decade: those include amplifiers [70], spin switches [101], polariton transistors [102], tunneling diodes [103], low power soliton transport [104], that were demonstrated at cryogenic temperatures based on GaAs microcavities. They rely on the design and fabrication of microcavities with specific geometries, playing with the dimensionality and the control of polariton potential. The development of building blocks for the photonic management in room-temperature operated microcavities is therefore crucial in order to confine the polaritons and to control their flow and the collection of their emission. The recent demonstration of GaN and ZnO microcavities on patterned mesas on Si substrates allows to control the in-plane geometry, and possibly the dimensionality, of their polariton condensates [65]. The polaritons can also be confined in 0D-traps [58]. Increasing the complexity of the photonic structures, the actual realization of a photonic crystal allowing a stronger control of the photonic bands [105,106] is still very challenging. Their equivalents with ZnO nanowire polariton lasers have however been successfully realized: a trap compatible with an evaporative cooling of the polariton condensate was recently demonstrated [107], as well as a periodic modulation of the polariton bands [52] and the coupling to ring resonators [86].

The organic and inorganic active layers both have interesting features and can be combined into a single microcavity, where the Frenkel excitons of the organic material and the Wannier excitons of the inorganic semiconductor are hybridized through the strong coupling with the confined photons, as proposed [59] and demonstrated [108] for III-V quantum wells and porphyrin molecules up to 100K. Such a scheme has recently been extended to room-temperature with an active layer combining ZnO and an organic molecule [109]. This opens the way to an easier polariton relaxation, and possibly electrical injection, into organic polariton lasers.

To conclude, we have reviewed the broad spectrum of organic and inorganic microcavities which excitons are robust enough to allow for the strong coupling regime at room temperature, and for the demonstration of polariton condensation with the most mature systems, i.e. GaN, ZnO, CuBr and some organic molecules. Each of them present strong specificities, related to the strength of the exciton binding and of the Rabi splitting, as well as the nature of the stimulated scattering process involved in the condensate formation. Their wide temperature range of condensation allows exploring the thermodynamics, and the frontier between the out-of-equilibrium and equilibrium regimes. As in GaAs and CdTe microcavities, the spatial dynamics and the spatial coherence of the condensate are impacted by the interplay between the repulsive interaction with the exciton reservoir, the ballistic propagation of the condensate and its localization due to the photonic disorder. The achievement of room-temperature polariton condensates has paved the way to the development of engineered polariton structures, where the photon confinement is managed in order to create periodic crystals, 1D polariton wires or 0D polariton traps.